# CinemaGazer: a System for Watching Videos at Very High Speed

Kazutaka Kurihara
National Institute of Advanced Industrial Science and Technology
Chuo Dai 2, 1-1-1 Umezono, Tsukuba, Ibaraki, Japan, 3058568.

qurihara@unryu.org

## ABSTRACT
This paper presents a technology that enables the watching of videos at very high speed. Subtitles are widely used in DVD movies, and provide useful supplemental information for understanding video contents. We propose a "two-level fast-forwarding" scheme for videos with subtitles, which controls the speed of playback depending on the context: very fast during segments without language, such as subtitles or speech, and "understandably fast" during segments with such language. This makes it possible to watch videos at a higher speed than usual while preserving the entertainment values of the contents. We also propose "centering" and "fading" features for the display of subtitles to reduce fatigue when watching high-speed video. We implement a versatile video encoder that enables movie viewing with two-level fast-forwarding on any mobile device by specifying the speed of playback, the reading rate, or the overall viewing time. The effectiveness of our proposed method was demonstrated in an evaluation study.

## Categories and Subject Descriptors
H5.1. Information interfaces and presentation (e.g., HCI): Multimedia Information Systems; H5.2. Information interfaces and presentation (e.g., HCI): User Interface.

## 1. INTRODUCTION
The amount of information on the Internet continues to increase. In addition to text-based information, there is an increasing diversity of other media types including audio-based (e.g., podcasts) and video-based information. For example, as many as 600 movies (with a total length of ~25 hours) are submitted to YouTube each minute [6]. In this context, there are two possible approaches to help users to efficiently find and review the information they need/want: reduce the amount of information to a manageable level by appropriate filtering, or improve users' ability to consume information quickly. The former approach has been thoroughly studied in the fields of searching, recommendation, and summarization of information, and innovations in this area have been widely applied. However, these technologies do not change factors related to information consumption, which remains the bottleneck in the process model. Hence, it is necessary to develop approaches that address this issue.

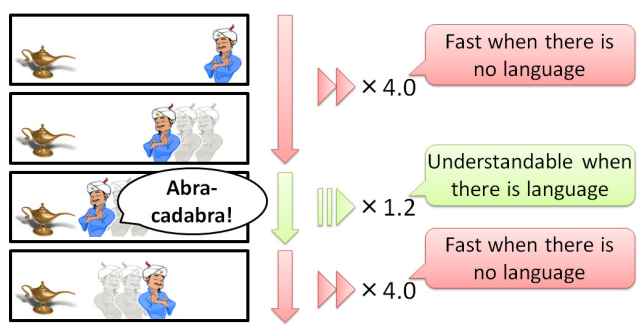

**Figure 1. System overview**

Here, we propose one such system. Many methods have been proposed for the quick review of text-based information [11]. Similarly, the review of audio and video information can be sped up by controlling playback speed using many standard media players, such as commercial video recorders, portable devices such as iPods, and PCs. However, only up to about twice normal speed is available using these players because the maximum playback speed at which users can process and effectively understand content linguistically has a certain upper limit.

As the first step to enable video viewing at high speed, we propose the "two-level fast-forwarding" method for watching videos with subtitles, which are widely used in commercial DVD movies [1]. This method exploits the difference between the maximum acceptable speeds of video with and without language. That is, it controls the speed of video playback depending on the context: very fast during segments without language, and understandably fast during segments with language. This makes it possible to watch video at a higher speed than when simply fast-forwarding, in a way that also enables a clearer linguistic understanding of the contents (Figure 1).

---
[1]In this paper we utilize "static" subtitles, which are not continuous and running like news feeds.

The rest of this paper is organized as follows. The next section reviews related work. Then the different methods for fast-forwarding video are introduced. After that, we describe our prototype system implementation in detail. Finally, we report the results of a study that we conducted to evaluate our method.

## 2. RERATED WORK

Many media players allow users to change the playback speed. Foulke et al. reported the SOLAFS algorithm for changing the speed of speech without pitch shifting, and concluded that it was effective for rapid understanding of content [3]. Vermi et al. proposed a method for improving listener comprehension of fast-forwarded speech by displaying text via speech recognition [19]. Aoki et al. proposed a fast-forwarding interface for reviewing music, using only auditory information [5]. Here, we consider the two modalities of audio and video simultaneously. In this context, the differences in the maximum acceptable speed between these modalities must be addressed.

Many studies have investigated issues related to changing the playback speed of videos. Kiyoyama et al. proposed a hardware device that generates low-speed speech output in combination with normal-speed video output from real-time video streams on television [13], which is effective mainly for elderly people who cannot keep up with the original speech streams. Cheng et al. [7] proposed "adaptive fast-forwarding," which enables adjustment of the current playback speed based on the complexity of the present scene and predefined semantic events, but they muted the audio and thus did not simultaneously manage that modality. Peker et al. [14][15] developed a method to accelerate playback speed according to visual and audio analyses in the video to maintain a "constant pace." We adopt a similar approach. However, we focus more on "language" modality, especially subtitles, which we believe is an important clue to allow users to watch videos much faster.

Displaying multiple information streams simultaneously is another approach to achieve quick or deep understanding of contents. There have been many reports about the use of three-dimensional localization of sound to enable searching and selection among multiple audio streams. Vazquez-Alvarez et al. discussed the psychological burden of this approach [18]. Forlines proposed a content aware video presentation on high-resolution displays [1], which renders multiple parts of a video content based on image-based shot/scene detection technologies to enhance viewing experiences. Fabro et al. [12] reported a tool for fast non-sequential hierarchical video browsing, which proposed parallel style views for a content. In addition, many types of software are available on the web that enable simultaneous viewing of multiple videos to reduce the total time required to watch them[2]. We limited the present research to the viewing of a single video stream, because we consider audio simultaneously, and because the multistream approach imposes too great a burden on most users. However, as our proposed method could easily be applied to a multistream approach, this may be a worthwhile future research direction.

In this study, we focus on techniques for watching a video from the beginning to the end sequentially, without skipping any parts.

Thus, interactive navigation techniques or summarization techniques for videos, such as those proposed previously [2][4][8][9][16][17][20], are outside the scope of this paper. However, highlights detection techniques for the interactive navigation or the summarization such as one reported by Diyakaran et al. [2] can be applicable to our system in a future to determine which parts of a movie should playback faster or slower.

## 3. FAST-FORWARDING OF VIDEOS

In this section, we discuss possible approaches for achieving understandable fast-forwarding of videos.

### 3.1 Various Fast-forwarding Methods

Video content, such as commercial DVD movies, has the following components:

1. Main video
2. Language 1: subtitles
3. Language 2: speech
4. Auxiliary audio

The main video is the original visual information sequence excluding the subtitles. The subtitles are textual information overlaid on the main video, and are provided as one of the language information streams. Speech is an audio information stream related to the human voice, to which each subtitle corresponds. The auxiliary audio is the other audio information stream excluding speech (e.g., background music and sound effects). There are several possible approaches for enabling fast-forwarding of videos taking into consideration synchronization between the main video and the language information channels (i.e., whether the start and end points of displayed language information correspond to the main video), and whether elimination of some parts of the language information should be allowed. Figure 2 shows the possible approaches with time series visualization with regard to the main video and language information (i.e., subtitles and/or speech).

The *Original* approach in the figure indicates normal video playback, whereas *High-speed (Simple)* shows simple fast-forwarding, which involves simply increasing the speed of playback. SOLAFS may be applied here to avoid changing the pitch of the audio information. This approach perfectly preserves synchronization between the main video and the language information. Many standard commercial video recorders have this simple fast-forwarding functionality enabling speeds up to ×1.3 or ×2.0. However, when much faster speeds are required, the speech or subtitles become difficult to follow using this approach.

*High-speed (Language Picking)* is an approach proposed by Seiyama et al. [13] that involves playing the main video at high speed, with some or all of the language information overlaid with reasonable speed conversion. As the speed of the main video is static, this method sacrifices synchronization between the main video and the language information to maintain an understandable speed of the latter. Thus, it results in marked time gaps between the main video and the language information if some language information cannot be eliminated and all of it is overlaid at an understandable speed.

---

[2]PluralMediaPlayer, MultiWindowMediaPlayer, WmpSzunl, etc.

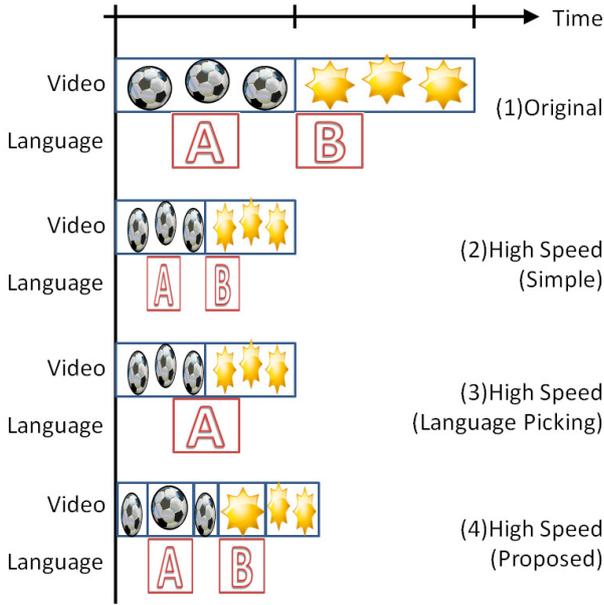

**Figure 2. Illustration of possible approaches for fast-forwarding of videos.**

The fourth method listed in Figure 2, *High-speed (Proposed)*, is our proposed two-level fast-forwarding method, which preserves perfect synchronization between the main video and the language information without eliminating any of the latter. This method involves controlling the speed of video playback depending on the context: very fast during segments without language (i.e., consisting of only the main video and the auxiliary audio) and understandably fast during segments with language (i.e., containing subtitles or speech).

Among the two available language information channels, we focus on subtitles because with appropriate training it is possible to achieve faster understanding of text than audio (i.e., speech). In addition, subtitles are widely available on commercial DVD movies, and the DVD format allows us to easily extract the timing and content information of subtitles. To utilize speech information when subtitles are not available would require preprocessing to extract the timing and content information of the speech. Future research will apply speech-recognition techniques to this preprocessing step.

### 3.2 Quantitative Analysis of Two-level Fast-forwarding

In this section, we formulate and analyze the various parameters of our proposed two-level fast-forwarding method and discuss its characteristics.

Here, we denote the time necessary to watch a video at the original speed as $L$ (sec). Video segments without language are played $S_m$ times faster than the original, and those with language are played $S_s$ times faster than the original. We also denote the ratio of the total durations of video with and without language as $r:1-r$. Then, we can formulate the actual time necessary $L_Q$ (sec) to watch the movie with two-level fast-forwarding as follows:

$$L_Q = \frac{rL}{S_m} + \frac{(1-r)L}{S_s}. \quad (1)$$

We can understand how two-level fast-forwarding can reduce viewing time by defining $L_Q/L$ as the compression ratio. In equation (1), $S_m$ and $S_s$, which are user-dependent parameters, contribute independently to $L_Q$. This suggests that with proper interfaces, it should be possible to develop independent training schemes for $S_m$ and $S_s$ allowing users to increase the watching speed. The ratio $r$ is the only movie-dependent parameter. We estimated the distribution of $r$ by sampling actual values from several movie DVDs. For this, we chose 15 DVD movies from various genres containing subtitles in Japanese: *Avatar*, *Climber's High*, *Death Note*, *Kaiji*, *Nankyoku Ryourinin*, *Pay it Forward*, *Ryusei no Kizuna*, *Die Dreigroschenoper*, *Saw*, *SP Yabouhen*, *The Thin Red Line*, *Moonstruck*, *Vantage Point*, *Yunagi no Machi Sakura no Kuni*, and *Doraemon: Nobita no Kekkon Zen'ya*. The average value of $r$ was 0.5777 and the standard deviation was 0.1184. This result suggests that about 50–70% of the total duration of these movies consists of video without language. Thus, on average, $S_m$ makes a greater contribution to the compression ratio than $S_s$ among the 15 movies chosen.

## 4. INTERFACES
### 4.1 Interfaces for Specifying Parameters for Two-level Fast-forwarding

In this section, we propose various interfaces for specifying the parameters for two-level fast-forwarding. Our proposed system has three interfaces as follows:

1. Specification of $S_m$ and $S_s$
2. Specification of $S_m$ and the user's reading rate
3. Specification of $L_Q$ and $S_s$

For two-level fast-forwarding, $L$ and $r$ are extracted from the video, and the system must know $S_m$ and $S_s$ to control the speed of playback. Point 1 above is the simplest and most direct interface for the system. Point 2 specifies the user's reading rate $S_t$ (characters/min) instead of $S_s$ [3]. In this case, $S_s$ is calculated for each subtitle depending on how much text information it contains. Then Equation 1 must be modified. When we denote $N_i$ (characters) as the total amount of text information in the $i$-th subtitle, we can formulate the actual duration $L_Q$ (sec) necessary to watch the movie with two-level fast-forwarding as follows:

$$L_Q = \frac{rL}{S_m} + \frac{60 \sum_i N_i}{S_t} \quad (2)$$

Here, we estimated the distribution of the required reading rate in the 15 movies mentioned above. The 10,913 subtitles for the 15 movies were processed by an optical character recognition system[4] and the total number of characters in the subtitles was determined. Dividing this number by the total duration of displayed subtitles yielded a required average reading rate of 329.9 characters/min with a standard deviation of 288.1 characters/min.

---

[3] The use of (words/min) would be more appropriate for subtitles in English.

[4] The subtitle contents in a movie DVD are usually embedded as independent image sequence.

As the normal Japanese reading rate is said to vary between 500 and 1000 characters/min [11] and some speed-reading training methodologies can apparently improve the rate to a level of 10,000 characters/min [11], average Japanese people can increase the compression ratio by using this interface and may achieve a greater increase with proper training.

Point 3 specifies the user's preferred $L_Q$. This is useful when the user has a specific limited time to watch a video. Given $L_Q$, the system calculates $S_m$ using equation (1) with minimal change to the specified $S_s$.

## 4.2 Interfaces for Displaying Subtitles

We implemented two interfaces with the aim of easing user burden when watching high-speed videos with subtitles.

### 4.2.1 Centering

Redundant gaze movements between the main video and subtitles are a problem when watching videos even with two-level fast-forwarding. When watching at the original speed, users can reserve enough time for gaze movements because the subtitles remain displayed for a sufficient period. During very high-speed watching, however, users may become tired due to frequent gaze movements or users may gaze at the subtitles only and quit watching the main video.

We applied the concept of "audience gaze localization" proposed by Kurihara et al. [10] in an attempt to minimize redundant gaze movements. Usually, the center of the main video is configured to be at the center of the display screen. On the other hand, subtitles are displayed at the screen periphery, usually at the bottom of the screen, to avoid disturbing the corresponding main video. Centering is a technique for displaying the centers of both at the same location, namely, moving the subtitles to the center of the screen.

We expect this technique to allow users to watch videos with their gaze fixed at the center of the screen and also gain sufficient understanding of the contents. On the other hand, centering may sometimes disturb users by occluding important video with subtitles. We revisit this issue in a later section.

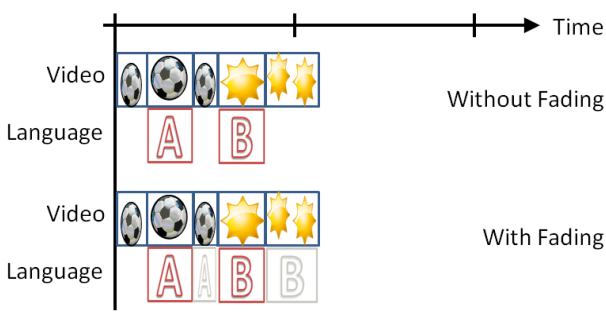

**Figure 3. Fading**

### 4.2.2 Fading

Fading is a technique where each subtitle remains displayed in translucent color until the next subtitle appears (Figure 3). We expect this technique to achieve better comfort when watching videos at high-speed because the duration of display of each subtitle is extended. We also expect that users will feel that synchronization between the main video and subtitles is preserved because the translucent visualization indicates that the subtitle has essentially expired. The extended display duration varies depending on the combination of $S_m$ and $S_s$. If these two parameters were assumed to be the same value, fading would take on average 2.623 times longer for the 10,913 subtitles in the 15 movies mentioned above.

## 5. SYSTEM IMPLEMENTATION

We implemented a two-level fast-forwarding video encoder as a prototype. We first attempted to implement a movie player prototype that changes the speed of playback in real-time, but it was difficult to control $S_m$ and $S_s$ precisely while keeping audio and video synchronized. The implemented encoder can instantly output target video files with any combination of $S_m$ and $S_s$. Consequently, we can realize similar functionality to a movie player prototype (i.e., $S_m$ and $S_s$ can be changed at any time during viewing, and the changes become effective immediately) using conventional media players by preparing video files of all of the possible combinations of $S_m$ and $S_s$ in advance.

The video encoder requires preprocessing to obtain a video with its subtitle information. When a commercial DVD movie is used as the source data, graphical user interface (GUI)-based DVD rippers, such as DVD Fab [5], can be used for this purpose. If a command line user interface (CUI) is preferred, the open-source tool VSRip is also available [6]. As the result of preprocessing, the standard format files for subtitle information (i.e., .idx and .sub files) are generated from the source DVD in addition to the video file itself.

The video encoder, which was implemented in C#, uses the preprocessed files as inputs and generates a two-level fast-forwarding video. Users specify the necessary system parameters and the preferred features of subtitles, as mentioned in the *Interfaces* section, via the GUI. The system generates AviSynth [7] script files that trim, stretch, shrink, and conjoin the source video based on the system parameters. Built-in functions of AviSynth, such as TimeStretch, AssumeFPS, and ChangeFPS, are used to stretch and shrink video and audio without shifting the pitch. The subtitles are overlaid as images using the Layer function. The AviSynth script files are simply text files, but are directly playable as videos with Windows Media Player. Furthermore, we implemented a functionality to convert the output AviSynth script files into conventional video formats such as MP4 and WMV, using FFmpeg [8] to allow them to be played on a wide variety of devices, such as smart phones.

## 6. EVALUATION

In this section, we report the results of an evaluation of two-level fast-forwarding in an example context. Ten volunteers (two men in their 30s, seven men in their 20s, and one woman in her 20s) participated in this study.

### 6.1 Method

All of the participants performed the following five tasks:

---

[5] http://www.dvdfab.com/

[6] http://sourceforge.net/projects/guliverkli/files/VSRip/

[7] http://avisynth.org/mediawiki/Main_Page

[8] http://ffmpeg.org/

1. Reading rate task
2. Speech-understanding-rate task
3. Subtitle-understanding-rate task
4. Main-video understanding-rate task
5. Two-level fast-forwarding with centered and faded subtitles task.

The order of execution of tasks 2, 3, and 4 was randomized for each participant to counterbalance order effects, but tasks 1 and 5 were always performed first and last, respectively. We performed task 1 always first because it was independent with respect to the other tasks. On the other hand, task 5 had to be performed always last because it used information obtained by tasks 3 and 4. A notebook PC (Sony VAIO VPCF1, with an LCD screen 36.4 cm in width and 30.0 cm in height) was placed on a desk. Each participant sat on a chair and maintained a normal distance from the PC's LCD screen suitable for standard PC use. The participants used headphones for audio, and set the volume according to their personal preference.

In task 1, part of the novel *Kokoro*, by Soseki Natsume, was displayed on the screen. The text consisted of 16 lines, each of which contained 30 characters. The necessary time to read the text was measured for each participant, and the reading rate (characters/min) was determined.

In tasks 2–5, the screen displayed a movie using Windows Media Player in full-screen mode. The title of the movie was *Doraemon: Nobita no Kekkon Zen'ya*, and a 7-min fragment from the beginning was encoded into MP4 files (video: VGA 8 bit AVC/h264 23.98 fps 1200 kb/sec, audio: AAC 48 kHz 2.0 ch 128 kb/sec) with various combinations of the system parameters, such as $S_m$ and $S_s$.

In task 2, the participants watched the movie without subtitles (i.e., the main video and all audio information) and were asked about the maximum playback speed at which they could understand the speech. The playback speed was varied from ×1.0 to ×10.0 in increments of ×0.5, and simple fast-forwarding (i.e., $S_m = S_s$) without pitch shifting was used.

In task 3, the participants watched the movie with subtitles (i.e., the main video, the subtitles, and all audio information), and were asked to focus on the subtitles only. They were asked about the maximum playback speed at which they could read and understand the subtitles. The playback speed was again varied from ×1.0 to ×10.0 in increments of ×0.5, and simple fast-forwarding without pitch shifting was used.

In task 4, the participants watched the movie with neither subtitles nor audio (i.e., the main video only), and were asked about the maximum playback speed at which they could understand the video context. Because the audio was muted, the participants were instructed that it was not necessary for them to try to guess the speech contents. The playback speed was again varied from ×1.0 to ×10.0 in increments of ×0.5, and simple fast-forwarding was used.

In task 5, the participants watched the movie with subtitles (i.e., the main video, the subtitles, and all audio information) using two-level fast-forwarding without pitch shifting, and were asked to subjectively evaluate their satisfaction. The playback speed was fixed for each participant as his/her upper limit: e.g., if a participant answered ×2.5 in task 3 and ×6.0 in task 4, $S_m$ was set as 6.0 and $S_s$ was set as 2.5. Each participant watched three movies under different conditions with regard to subtitles in random order: the original condition denoted as *O* displayed subtitles at the original position (i.e., at the bottom of the screen), the centering condition denoted as *C*, and the centering and fading condition denoted as *CF*. The subjective evaluations of satisfaction were indicated as real values between 1 and 5 where 1 was "very uncomfortable," 3 was "neutral," and 5 was "very comfortable." Furthermore, in this task, the eye gaze points on the screen were measured in two of the ten participants using a Tobii X60 eye tracker at 60 Hz.

After performing the abovementioned five tasks, the participants were interviewed for any feedback and comments regarding the tasks.

## 6.2 Results

Tasks 1 and 3 revealed a correlation between reading rate and the maximum playback speed at which one could read and understand the text of the subtitles. Figure 4 shows a scatter plot of the data. The correlation coefficient was 0.1426, which was considerably low.

From tasks 2–4, we determined the ratios of the number of participants who could subjectively understand the contents with regard to the various playback speeds of speech, subtitles, and main video (Figure 5). The solid lines in Figure 5 are the results of 2-parameter logistic regression analysis using the least squares method: $y = 1/(1 + exp(a(x-b)))$. In task 4, seven of the ten participants responded that even the ×10.0 speed was not at their upper limit for understanding.

From task 5, we obtained the satisfaction levels with respect to the three subtitle display conditions when watching videos at very high speed using two-level fast-forwarding, as shown in Figure 6. The average values of satisfaction for the *O*, *C*, and *CF* conditions were 1.95, 3.43, and 3.75, respectively. The standard deviations were 0.599, 0.778, and 0.677, respectively.

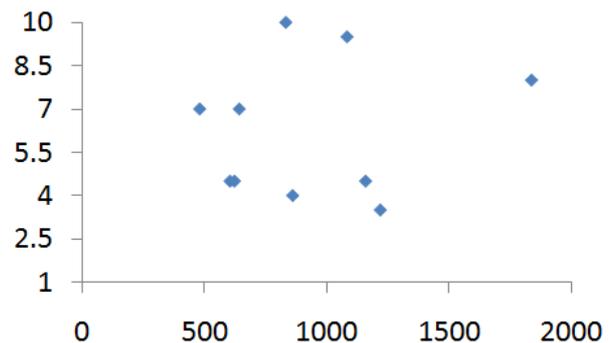

**Figure 4. Correlation between reading rate (characters/min, x-axis) and maximum playback speed of subtitles (y-axis).**

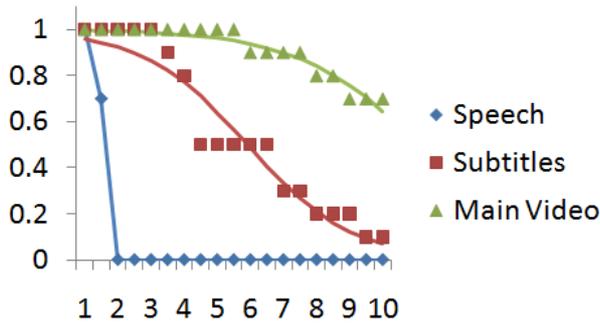

**Figure 5. Ratios of number of participants who could subjectively understand the contents (y-axis) with regard to the various playback speeds (x-axis).**

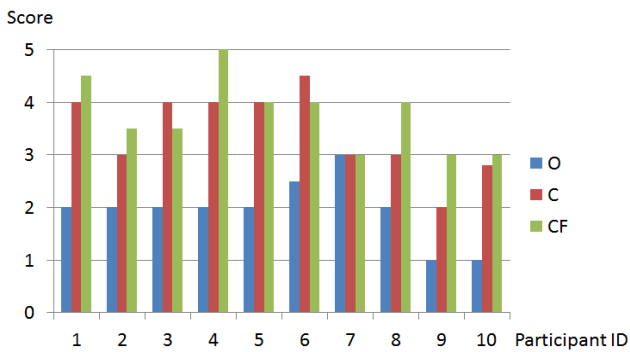

**Figure 6. Participants' satisfaction levels with respect to the three subtitle display conditions.**

## 6.3 Discussion

The low correlation shown in Figure 4 implies that the required skills for reading normal text and reading subtitles in a video differ. This difference may be derived from the fact that the timing and amount of subtitles displayed on the screen are not predictable in advance when reading subtitles. On the other hand, when reading normal text, these can be predicted in advance and efficient eye movements can be planned. Our implemented prototype can set the speed of playback of video with language depending on the user's reading rate $S_t$ (characters/min), but it is necessary to evaluate the effectiveness of this functionality in future studies.

The data shown in Figure 5 indicate that users may subjectively understand subtitles faster than speech, and subjectively understand main video faster than the two other types of language information. Individual differences were smallest in the speech-understanding rate and largest in the main video-understanding rate. If we assume that parameter $b$ (the playback speed at which the ratio of the number of participants who can subjectively understand the contents is 0.5) is a representative value of the understandable playback speeds, then these speeds would be ×1.552 for speech, ×5.910 for subtitles, and ×11.14 for the main video. Thus, two-level fast-forwarding with equation (1) would result in an average 85.51% reduction in viewing time by abandoning speech information.

However, this result was obtained because of the characteristics of the video used in this study. First, the total duration of the video was short enough to prevent the participants from feeling fatigue when watching at high speed[9]. In addition, the video was an animation for children, and therefore the speed of scene changes and camera angle movements may be small overall, and the speed of the speech may be slow. Therefore, this result may represent close to the best performance of our two-level fast-forwarding system. Different results will likely be obtained for longer movies, action movies, or movies that contain faster or more complex dialog. That said, another informal study in our laboratory on such movies also showed that subtitle understanding was faster than speech understanding. One direction of future research will be to combine the two-level fast-forwarding method with adaptive fast-forwarding [7] that dynamically adjusts the playback speed of main video to make them understandable based on the complexity of the present scene using a vision-based technique.

We also have to take into consideration that the obtained result was not based on participants' objective content understanding but subjective one. We believe the individual subjective understanding is the appropriate evaluation criterion for satisfactory consumptions of contents such as entertainment movies. However, objective understanding evaluations will also be needed when the users should learn something from contents such as news or lecture videos.

Finally, the participants' satisfaction levels under the $O$, $C$, and $CF$ conditions shown in Figure 6 indicate that the combination of two-level fast-forwarding and centering was relatively effective for watching videos at high speed, because the $O$ condition was not more satisfactory than the other two conditions for all the participants. This result was also supported by the comments from multiple participants, who reported that it was uncomfortable because they had to move their eye gaze frequently between the subtitles and the video contents. On the other hand, we can infer that under the $C$ and $CF$ conditions (i.e., the subtitles were displayed at the center of the screen), the dissatisfaction of participants was reduced because they could fix their eye gaze around the center of the screen.

To confirm this inference, we visualized the gaze of one participant (ID: 10) on the screen during task 5 under both the $O$ and $CF$ conditions. Figure 7 and 8 shows a scatter plot for each condition and Figure 9 shows a histogram of relative frequency of gaze duration with regard to the $y$-coordinate. These figures indicate that the gaze was focused mainly around the center of the screen ($y = 600$) in the $CF$ condition, with regard to the $y$-coordinate. On the other hand, the gaze was spread around the bottom ($y = 350$) and around the center of the screen ($y = 600$) in the $O$ condition. These observations indicate that in the $CF$ condition the "audience gaze localization" [10] was achieved and that it was effective.

There were no large differences in the satisfaction levels between the $C$ and $CF$ conditions. From the participants' comments, this appeared to be because the extra display time of subtitles realized by fading was nearly zero when the dialogs were too fast to have sufficient non-language parts between subtitles. It may also have been because the faded subtitle content did not fit semantically with the main video content when segments without language were too long. Although the former problem is not solvable in principle, the latter problem could be solved by setting a maximum duration for a fading subtitle, after which it disappears.

---

[9] However, the duration is only 224 sec when one watches this video to the end with $S_m = 11.14$ and $S_s = 5.910$.

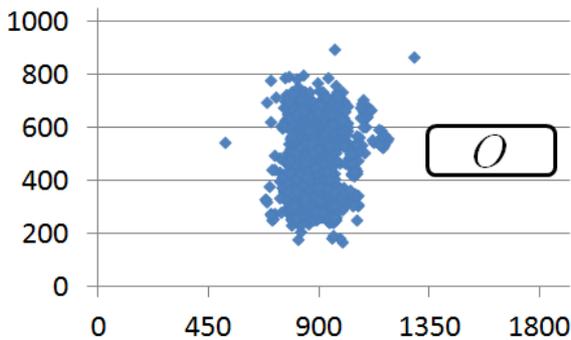

**Figure 7. Scatter plot of eye gaze with the *O* condition on the 1920×1080 screen. The horizontal axis shows the *x*-coordinate (pixels), and the vertical axis shows the *y*-coordinate (pixels).**

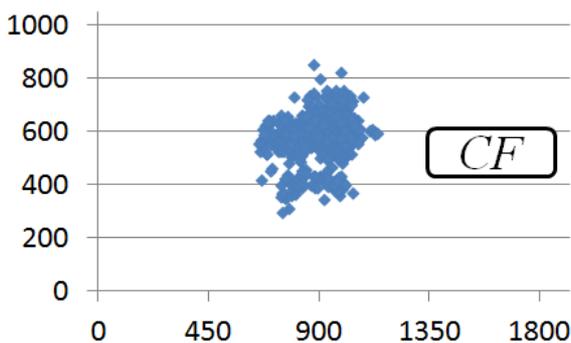

**Figure 8. Scatter plot of eye gaze with the *CF* condition on the 1920×1080 screen. The horizontal axis shows the *x*-coordinate (pixels), and the vertical axis shows the *y*-coordinate (pixels).**

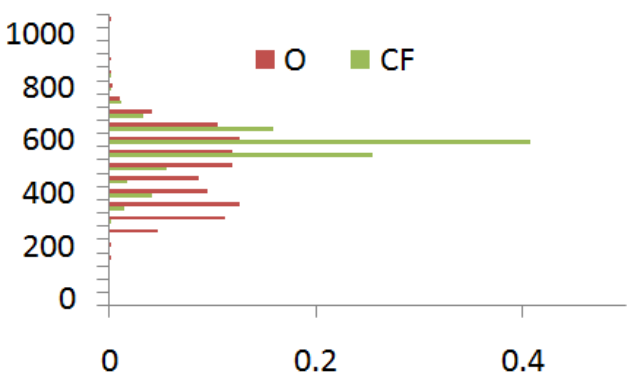

**Figure 9. Histogram of relative frequency of gaze duration with regard to the *y*-coordinate on the screen.**

## 7. CONCLUSION

We proposed a two-level fast-forwarding system for independently controlling the playback speed of video without language and video with language to enable very high-speed viewing. This technology was implemented as a video encoder for commercial DVD movies containing subtitles. The effectiveness of a combination of two-level fast-forwarding and a centering interface in a context was demonstrated, and an example that enabled an average 85.51% reduction in the original watching duration was described. In future research, it will be important to estimate the upper limits of $S_m$ and $S_s$ for various video genres. Another promising area of research will involve the introduction of audio processing and speech recognition techniques to enable the two-level fast-forwarding of videos without subtitle information.

## 8. ACKNOWLEDGMENTS

This work was partially supported by JSPS KAKENHI 23700155.